\documentclass[aps,preprint]{revtex4}%
\usepackage{amsfonts}
\usepackage{amsmath}
\usepackage{amssymb}
\usepackage{graphicx}%
\begin{document}
\title{Experimental Evidence for a Dynamical Non-Locality Induced Effect in Quantum
Interference Using Weak Values}
\author{S. E. Spence and A. D. Parks }
\affiliation{Quantum Processing Group, Electromagnetic and Sensor Systems Department, Naval
Surface Warfare Center, Dahlgren, VA 22448 USA}

\begin{abstract}
The quantum theoretical concepts of modular momentum and dynamical
non-locality, which were introduced four decades ago, have recently been used
to explain single particle quantum interference phenomena. Although the
non-local exchange of modular momentum associated with such phenomena cannot
be directly observed, it has been suggested that effects induced by this
exchange can be measured experimentally using weak measurements of pre- and
post-selected ensembles of particles. This paper reports\ on such an optical
experiment that yielded measured weak values that were consistent with the
theoretical prediction of an effect induced by a non-local exchange of modular momentum.

\end{abstract}
\date{Experimental / Draft Completion......January 19, / March 12, 2010 \ \ }
\startpage{1}
\endpage{102}
\maketitle

\section{Introduction}

Because it differs fundamentally from the interference phenomena of classical
physics, quantum interference has remained a continuing topic for discussion
and debate since quantum theory's early days. The essence of this difference
is exhibited by the two-slit experiment. From both the classical and
Schr\"{o}dinger wave perspectives, the two slit interference pattern is easily
described in terms of the overlapping contributions of the wave which have
passed through each slit. The wave perspective also explains the disappearance
of the interference pattern when one of the slits is closed.

However, interference experiments using low intensity electron or photon beams
in which only one particle at a time passes through a two-slit apparatus have
shown that the accumulated effect when both slits are open is an interference
pattern like that produced by higher intensity ensembles and that the pattern
likewise disappears when one slit is closed, e.g. \cite{Ton}. This peculiar
behavior necessitates an answer to the question "how does a particle passing
through one slit sense that the other slit is open or closed?" when
interference is considered from the perspective of a single quantum particle.

Although this question concerning single particle behaviour has been answered
and explained theoretically in terms of a non-local exchange of modular
momentum \cite{APP1,APP2}, there have been no direct experimental observations
of such an exchange to support this explanation. This lack of observations is
due to the fact that the conditions required to observe a non-local exchange
of modular momentum are precisely those that make the associated modular
variable completely uncertain and unobservable. Recently, however, it was
suggested that an experimental methodology using weak measurements performed
on a pre- and post-selected ensemble of particles could be exploited in order
to observe an effect \emph{induced} by a non-local exchange of modular
momentum. This methodology was illustrated by a \textit{gedanken} experiment
which used a twin Mach-Zehnder interferometer to duplicate relevant aspects of
the two-slit interference experiment \cite{TACKN}.

This paper reports the results of an optical twin Mach-Zehnder interferometer
experiment similar to that described in the above \textit{gedanken}
experiment. This experiment yielded measured weak values that were consistent
with the associated theoretical prediction describing the effect induced by a
non-local exchange of modular momentum. The remainder of this paper is
organized as follows: in the next section the theories of modular momentum,
dynamical non-locality, weak measurements, and weak values are briefly
summarized. A description of the experimental apparatus and an overview of the
experiment are presented in section III. The experimental results are
discussed in section IV. Concluding remarks comprise the final section of this paper.

\section{Summary of the Theories}

\subsection{Modular Momentum and Dynamical Non-locality}

Consider a quantum particle propagating in the positive $y$-direction
perpendicular to the plane of two symmetric slits which are separated by a
distance $\ell$ in the $x$-direction (the slit at $x-\ell$ will be referred to
as the left slit). At time $t$ after the particle passes through the slits its
wavefunction is the superposition
\begin{equation}
\psi\left(  x,y,z,t\right)  =\frac{1}{\sqrt{2}}\left\{  \varphi\left(
x-\ell,y,z,t\right)  +e^{i\alpha}\varphi\left(  x,y,z,t\right)  \right\}  ,
\label{1}%
\end{equation}
where the $\varphi$'s are assumed to be identical "wave packets" which do not
overlap at $t=0$ and $\alpha$ is their relative phase. Although information
about $\alpha$ can be obtained from the spatial interference pattern
$\left\vert \psi\left(  x,d,z,\tau\right)  \right\vert ^{2}$ produced by an
ensemble of such particles on a screen parallel to and at an appropriate
distance $d$ from the plane of the slits at time $\tau>0$, there are no local
measurements using operators of the form $\widehat{x}^{j}\widehat{p}_{x}^{k}$,
where $j$ and $k$ are non-negative integers, that can be performed upon the
initial non-overlapping wave packets that will determine $\alpha$. The
relative phase $\alpha$ is thus a non-local feature of quantum mechanics.

The induced momentum uncertainty and the Heisenberg uncertainty principle are
traditionally used to explain the loss of the interference pattern when one
slit is closed. However, measuring which slit the particle passes through does
not necessarily increase the momentum uncertainty. This - along with the fact
that position and momentum observables and their moments are not sensitive to
relative phase (prior to wave packet overlap) - suggests that these
observables, as well as the Heisenberg uncertainty principle, are not the
appropriate physical concepts for describing quantum interference phenomena.

The (modular) operator $e^{-\frac{i}{\hbar}\widehat{p}_{x}\ell}$ and its
modular property, however, do provide an alternative physical basis for the
rational description of quantum interference. Unlike the operators
$\widehat{x}^{j}\widehat{p}_{x}^{k}$, the expectation value of the operator
$e^{-\frac{i}{\hbar}\widehat{p}_{x}\ell}$ with respect to $\psi\left(
x,y,z,t\right)  $ is sensitive to $\alpha$ - even when the two\ wavepackets
don't overlap. This sensitivity results from the action of $e^{-\frac{i}%
{\hbar}\widehat{p}_{x}\ell}$ upon $\varphi\left(  x,y,z,t\right)  $ which
overlaps the two wavepackets in eq. (\ref{1}) by translating $\varphi\left(
x,y,z,t\right)  $ to $\varphi\left(  x-\ell,y,z,t\right)  $. Also, since
$e^{-\frac{i}{\hbar}\widehat{p}_{x}\ell}$ is invariant under the replacement
$\widehat{p}_{x}\rightarrow\widehat{p}_{x}-n\frac{h}{\ell}$, $n=0,\pm
1,\pm2,\cdots$, (because $e^{-\frac{i}{\hbar}\left(  -n\frac{h}{\ell}\right)
\ell}=e^{2in\pi}=1$), it depends upon values of the \emph{modular momentum}
$p_{x}\operatorname{mod}\left(  \frac{nh}{\ell}\right)  \equiv
p_{x,\operatorname{mod}}\in I\equiv\lbrack0,\frac{h}{\ell})$ instead of those
of $p_{x}$. This modular property establishes a fundamental relationship
between modular momentum uncertainty and quantum interference via the complete
uncertainty principle: "$\widehat{p}_{x,\operatorname{mod}}$ is completely
uncertain (i.e. all its values are uniformly distributed over $I$) if and only
if $\left\langle e^{-\frac{i}{\hbar}n\widehat{p}_{x}\frac{\ell}{h}%
}\right\rangle =0$ for every positive integer $n$". When this principle is
applied to the two slit case, it is found that while the required expectation
value with respect to $\psi\left(  x,y,z,t\right)  $ does not vanish for
$n=1$, it does vanish for every $n$ when the expectation value is with respect
to $\varphi\left(  x,y,z,t\right)  $. Thus, when the left slit is closed, i.e.
it is known that the particle passed through the right slit, then $\widehat
{p}_{x,\operatorname{mod}}$ becomes completely uncertain so that all knowledge
about $p_{x,\operatorname{mod}}$ is lost and the interference pattern vanishes.

The Heisenberg equation of motion provides the formalism for describing and
understanding the notion of \emph{dynamical non-locality}. Within the context
of two slits, the Heisenberg equation of motion for $e^{-\frac{i}{\hbar
}\widehat{p}_{x}\ell}$ is given by
\begin{equation}
\frac{d}{dt}e^{-\frac{i}{\hbar}\widehat{p}_{x}\ell}=\frac{i}{\hbar}\left[
\widehat{H},e^{-\frac{i}{\hbar}\widehat{p}_{x}\ell}\right]  =\frac{i}{\hbar
}\left(  \widehat{V}\left(  x\right)  -\widehat{V}\left(  x-\ell\right)
\right)  e^{-\frac{i}{\hbar}\widehat{p}_{x}\ell}, \label{1a}%
\end{equation}
where $\widehat{H}=\frac{\widehat{p}_{x}^{2}}{2m}+\widehat{V}\left(  x\right)
$ is the system Hamiltonian and $\widehat{V}\left(  x\right)  $ ( $\widehat
{V}\left(  x-\ell\right)  $ ) is the potential operator for the right (left)
slit. This is a non-local equation of motion and therefore has no classical
analogue: only the potential at each slit is involved in the rate of change of
$e^{-\frac{i}{\hbar}\widehat{p}_{x}\ell}$ - i.e. there are no forces involved
- and the potential at the left slit influences this rate of change even if
$\psi\left(  x,y,z,0\right)  =\varphi\left(  x,y,z,0\right)  $ - i.e. when the
particle is initially localized at the right slit. Consequently, the effect of
closing the left slit produces non-locally a change in modular momentum while
leaving the expectation values of the associated moments of momentum
unchanged. More specifically, the modular operator is conserved when both
slits are open since $\widehat{V}\left(  x\right)  =\widehat{V}\left(
x-\ell\right)  $ so that $\frac{d}{dt}e^{-\frac{i}{\hbar}\widehat{p}_{x}\ell
}=0$. However, if the left slit is closed and the particle is localized at the
right slit, then $\left(  \widehat{V}\left(  x\right)  -\widehat{V}\left(
x-\ell\right)  \right)  \neq0$ $\neq\frac{d}{dt}e^{-\frac{i}{\hbar}\widehat
{p}_{x}\ell}$ and the modular momentum is changed non-locally as a result of
the change in the potential at the left slit. Interference is destroyed and
the modular momentum becomes completely uncertain - thereby rendering it
unobservable. In fact, it is in this manner that the complete uncertainty
principle also reconciles dynamical non-locality with causality.

For additional details concerning the theory of modular momentum and dynamical
non-locality the reader is invited to consult references
\cite{APP1,APP2,TACKN,AR}.

\subsection{Weak Measurements and Weak Values}

Although the exchange of modular momentum is not directly observable, it has
been suggested that dynamical non-locality induces effects which can be
observed using weak measurements of pre- and post-selected ensembles of
particles. Weak measurements arise in the von Neumann description of a quantum
measurement at time $t_{0}$ of a time-independent observable $\widehat{A}$
that describes a quantum system in an initial fixed pre-selected state
$\left\vert \psi_{i}\right\rangle =\sum_{J}c_{j}\left\vert a_{j}\right\rangle
$ at $t_{0}$, where the set $J$ indexes the eigenstates $\left\vert
a_{j}\right\rangle $ of $\widehat{A}$. In this description the Hamiltonian for
the interaction between the measurement apparatus and the quantum system is%
\[
\widehat{H}=\gamma(t)\widehat{A}\widehat{p}.
\]
Here $\gamma\left(  t\right)  =\gamma\delta\left(  t-t_{0}\right)  $ defines
the strength of the impulsive measurement interaction at $t_{0}$ and
$\widehat{p}$ is the momentum operator for the pointer of the measurement
apparatus which is in the initial state $\left\vert \phi\right\rangle $. Let
$\widehat{q}$ be the pointer's position operator that is conjugate to
$\widehat{p}$ and assume that $\left\langle q\right\vert \left.
\phi\right\rangle \equiv\phi\left(  q\right)  $ is real valued with
$\left\langle q\right\rangle \equiv\left\langle \phi\right\vert \widehat
{q}\left\vert \phi\right\rangle =0$.

Prior to the measurement the pre-selected system and the pointer are in the
tensor product state $\left\vert \psi_{i}\right\rangle \left\vert
\phi\right\rangle $. Immediately following the measurement the combined system
is in the state
\[
\left\vert \Phi\right\rangle =e^{-\frac{i}{\hbar}\int\widehat{H}dt}\left\vert
\psi_{i}\right\rangle \left\vert \phi\right\rangle =%
{\displaystyle\sum\nolimits_{J}}
c_{j}e^{-\frac{i}{\hbar}\gamma a_{j}\widehat{p}}\left\vert a_{j}\right\rangle
\left\vert \phi\right\rangle ,
\]
where use has been made of the fact that $\int\widehat{H}dt=\gamma\widehat
{A}\widehat{p}$. The exponential factor in this equation is the translation
operator $\widehat{S}\left(  \gamma a_{j}\right)  $ for $\left\vert
\phi\right\rangle $ in its $q$-representation. It is defined by the action
$\left\langle q\right\vert \widehat{S}\left(  \gamma a_{j}\right)  \left\vert
\phi\right\rangle =\left\langle q-\gamma a_{j}\right\vert \left.
\phi\right\rangle \equiv\phi\left(  q-\gamma a_{j}\right)  $ which translates
the pointer's wavefunction over a distance $\gamma a_{j}$ parallel to the
$q$-axis. The $q$-representation of the combined system and pointer state is%
\[
\left\langle q\right\vert \left.  \Phi\right\rangle =%
{\displaystyle\sum\nolimits_{J}}
c_{j}\left\langle q\right\vert \widehat{S}\left(  \gamma a_{j}\right)
\left\vert \phi\right\rangle \left\vert a_{j}\right\rangle .
\]

When the measurement interaction is strong, the quantum system is appreciably
disturbed and its state "collapses" to an eigenstate $\left\vert
a_{n}\right\rangle $ leaving the pointer in the state $\left\langle
q\right\vert \widehat{S}\left(  \gamma a_{n}\right)  \left\vert \phi
\right\rangle $ with probability $\left\vert c_{n}\right\vert ^{2}$. Strong
measurements of an ensemble of identically prepared systems yield
$\gamma\left\langle A\right\rangle \equiv\gamma\left\langle \psi
_{i}\right\vert \widehat{A}\left\vert \psi_{i}\right\rangle $ as the centroid
of the pointer probability distribution%
\begin{equation}
\left\vert \left\langle q\right\vert \left.  \Phi\right\rangle \right\vert
^{2}=%
{\displaystyle\sum\nolimits_{J}}
\left\vert c_{j}\right\vert ^{2}\left\vert \left\langle q\right\vert
\widehat{S}\left(  \gamma a_{j}\right)  \left\vert \phi\right\rangle
\right\vert ^{2} \label{2}%
\end{equation}
with $\left\langle A\right\rangle $ as the measured value of $\widehat{A}$.

A \emph{weak measurement} of $\widehat{A}$ occurs when the interaction
strength $\gamma$ is sufficiently small so that the system is essentially
undisturbed and the uncertainty $\Delta q$ is much larger than $\widehat{A}$'s
eigenvalue separation. In this case, eq.(\ref{2}) is the superposition of
broad overlapping $\left\vert \left\langle q\right\vert \widehat{S}\left(
\gamma a_{j}\right)  \left\vert \phi\right\rangle \right\vert ^{2}$ terms.
Although a single measurement provides little information about $\widehat{A}$,
many repetitions allow the centroid of eq.(\ref{2}) to be determined to any
desired accuracy.

If a system state is post-selected after a weak measurement is performed, then
the resulting pointer state is%
\[
\left\vert \Psi\right\rangle =\left\langle \psi_{f}\right\vert \left.
\Phi\right\rangle =%
{\displaystyle\sum\nolimits_{J}}
c_{j}^{\prime\ast}c_{j}\widehat{S}\left(  \gamma a_{j}\right)  \left\vert
\phi\right\rangle ,
\]
where $\left\vert \psi_{f}\right\rangle =%
{\displaystyle\sum\nolimits_{J}}
c_{j}^{\prime}\left\vert a_{j}\right\rangle $, $\left\langle \psi
_{f}\right\vert \left.  \psi_{i}\right\rangle \neq0$, is the post-selected
state at $t_{0}$. Since%
\[
\widehat{S}\left(  \gamma a_{j}\right)  =\sum_{m=0}^{\infty}\frac{\left[
-i\gamma a_{j}\widehat{p}/\hbar\right]  ^{m}}{m!},
\]
then%
\[
\left\vert \Psi\right\rangle =%
{\displaystyle\sum\nolimits_{J}}
c_{j}^{\prime\ast}c_{j}\left\{  1-\frac{i}{\hbar}\gamma A_{w}\widehat{p}%
+\sum_{m=2}^{\infty}\frac{\left[  -i\gamma\widehat{p}/\hbar\right]  ^{m}}%
{m!}\left(  A^{m}\right)  _{w}\right\}  \left\vert \phi\right\rangle
\approx\left\{
{\displaystyle\sum\nolimits_{J}}
c_{j}^{\prime\ast}c_{j}\right\}  e^{-\frac{i}{\hbar}\gamma A_{w}\widehat{p}%
}\left\vert \phi\right\rangle
\]
in which case%
\begin{equation}
\left\vert \Psi\right\rangle \approx\left\langle \psi_{f}\right\vert \left.
\psi_{i}\right\rangle \widehat{S}\left(  \gamma A_{w}\right)  \left\vert
\phi\right\rangle \label{2b}%
\end{equation}
so that%
\[
\left\vert \left\langle q\right\vert \left.  \Psi\right\rangle \right\vert
^{2}\approx\left\vert \left\langle \psi_{f}\right\vert \left.  \psi
_{i}\right\rangle \right\vert ^{2}\left\vert \left\langle q\right\vert
\widehat{S}\left(  \gamma\operatorname{Re}A_{w}\right)  \left\vert
\phi\right\rangle \right\vert ^{2}%
\]
or%
\begin{equation}
\left\vert \Psi\left(  q\right)  \right\vert ^{2}\approx\left\vert
\left\langle \psi_{f}\right\vert \left.  \psi_{i}\right\rangle \right\vert
^{2}\left\vert \phi\left(  q-\gamma\operatorname{Re}A_{w}\right)  \right\vert
^{2}. \label{3}%
\end{equation}
Here%
\[
\left(  A^{m}\right)  _{w}=\frac{%
{\displaystyle\sum\nolimits_{J}}
c_{j}^{\prime\ast}c_{j}a_{j}^{m}}{%
{\displaystyle\sum\nolimits_{J}}
c_{j}^{\prime\ast}c_{j}}=\frac{\left\langle \psi_{f}\right\vert \widehat
{A}^{m}\left\vert \psi_{i}\right\rangle }{\left\langle \psi_{f}\right\vert
\left.  \psi_{i}\right\rangle },
\]
with the \emph{weak value} $A_{w}$ of $\widehat{A}$ defined by
\begin{equation}
A_{w}\equiv\left(  A^{1}\right)  _{w}=\frac{\left\langle \psi_{f}\right\vert
\widehat{A}\left\vert \psi_{i}\right\rangle }{\left\langle \psi_{f}\right\vert
\left.  \psi_{i}\right\rangle }. \label{4}%
\end{equation}
From this expression it is obvious that $A_{w}$ is - in general - a complex
valued quantity that can be calculated directly from theory. Since
$\phi\left(  q\right)  $ is real valued, then eq.(\ref{3}) corresponds to a
broad pointer position distribution with a single peak at $\left\langle
q\right\rangle =\gamma\operatorname{Re}A_{w}$ with $\operatorname{Re}A_{w}$ as
the measured value of $\widehat{A}$. This condition occurs when both of the
following inequalities relating $\gamma$ and the pointer momentum uncertainty
$\Delta p$ are satisfied \cite{DSS,PCS} :%
\begin{equation}
\Delta p\ll\frac{\hbar}{\gamma}\left\vert A_{w}\right\vert ^{-1}\text{ and
}\Delta p\ll\underset{(m=2,3,\cdots)}{\min}\frac{\hbar}{\gamma}\left\vert
\frac{A_{w}}{\left(  A^{m}\right)  _{w}}\right\vert ^{\frac{1}{m-1}}.
\label{5}%
\end{equation}

It is important to keep in mind that although the weak measurement of
$\widehat{A}$ occurs at time $t_{0}$ so that $\left\vert \psi_{i}\right\rangle
$ and $\left\vert \psi_{f}\right\rangle $ are states at $t_{0}$, these states
result from states that are pre-selected and post-selected at times
$t_{i}<t_{0}$ and $t_{f}>t_{0}$, respectively. Therefore it is necessary to
propagate the pre-selected state forward in time from $t_{i}$ to $t_{0}$ and
the post-selected state backward in time from $t_{f}$ to $t_{0}$ in order to
calculate $A_{w}$ at $t_{0}$.

The reader is invited to consult references
\cite{TACKN,AR,AAV,DSS,AV,RSH,PCS,HK,DSJH} for additional details concerning
the theoretical and experimental aspects of weak measurements and weak values.

\section{The Experiment}

\subsection{Apparatus}

As mentioned above, the setup for this experiment follows that of the optical
\textit{gedanken} experiment discussed in \cite{TACKN} where a twin
Mach-Zehnder interferometer is used to replicate aspects of the two-slit
interference experiment. A schematic of the apparatus used in this experiment
is shown in figure 1. Here the paths followed by photons have been\ labeled
using the traditional "right" ($R$) and "left" ($L$) notation $R1,R2,\cdots
,R6,L2,L3,\cdots,L6$. For future reference an overlay of the "metaphorical"
two slits emulated by the twin Mach-Zehnder interferometer is also provided in
this figure. Note that paths $R4$ and $L4$ correspond to photon paths through
the right and left slits, respectively. Thus, blocking path $L4$ corresponds
to closing the left slit.

Since photons do not interact with one another, it is not necessary to perform
the experiment in such a manner that only one photon at a time traverses the
interferometer. Accordingly, large ensembles of photons of wavelength $637.2$
$nm$ produced by a classically intense laser diode source were used in this
experiment. A $150$ $\mu m$ diameter pinhole spatially filtered the photon
beam into a smooth Gaussian-like shape. The exiting beam had an optical power
of $24.5$ $\mu W$ ($\sim7.9\times10^{13}$ photons/s) and was collimated with a
$200$ $mm$ focal length lens. A mirror launched the collimated beam into the
interferometer via the input path $R1$. Three identical non-polarizing cube
50/50 beam-splitters - labeled BS1, BS2, BS3 in figure 1 - along with four
identical mirrors - labeled M1, M2, M3, M4 in figure 1 - formed the basic
architecture of the interferometer (the collection BS1, M1, M2, and BS2 (BS2,
M3, M4, and BS3) is hereafter referred to as "\emph{the first (second)
Mach-Zehnder}"). The beam emerging along path $R6$ was neutral density
filtered before reaching a $640\times480$ pixel resolution machine vision
camera which recorded the beam's two dimensional intensity distribution. The
optical power of the beam reaching the camera was approximately four to five
orders of magnitude smaller than that exiting the pinhole. Each camera pixel
had a size $7.4$ $\mu m$ $\times$ $7.4$ $\mu m$ and a $0-255$ digital
intensity range. The pixel saturation level exceeded the measured maximum
pixel intensity level of the images obtained from this experiment.

The \textit{gedanken} experiment utilized slightly tilted thin glass plates
placed at locations in paths $R2$ and $L2$ to perform weak measurements of the
projection operators $\left\vert R2\right\rangle \left\langle R2\right\vert $
and $\left\vert L2\right\rangle \left\langle L2\right\vert $ by producing
transverse spatial shifts in the photon paths that were small relative to the
uncertainty in the transverse position of a photon. The theoretically
predicted change in the weak values of these operators when path $L4$ is
blocked was interpreted as an observable effect induced in the first
Mach-Zehnder by an associated non-local exchange of modular momentum produced
by blocking path $L4$ in the second Mach-Zehnder (direct measurement of the
modular momentum exchange is not possible because blocking path $L4$ in the
second Mach-Zehnder makes the modular variable completely uncertain - thereby
destroying all information about the modular momentum).

In this experiment, however, a piezoelectrically driven computer controlled
stage was used instead to produce small changes in the location of mirror M1
(in the direction shown in figure 1) in order to produce a series of
transverse spatial shifts in the photon beam that could be made small compared
to the uncertainty in a photon's transverse position. This approach proved
more efficient than the tilted plate method and was equivalent to performing
weak measurements of the projection operator $\left\vert L2\right\rangle
\left\langle L2\right\vert $ located in path $L2$. As shown - both
theoretically and experimentally - below, the weak value of $\left\vert
L2\right\rangle \left\langle L2\right\vert $ changes in accordance with the
\textit{gedanken} experiment when path $L4$ is blocked. This change can also
be interpreted as a dynamical non-locality induced effect.

By avoiding the use of micro-positioners as much as possible, the setup was
passively stable for several tens of minutes. The entire apparatus was also
enclosed in a $1$ $m$ $\times$ $1$ $m$ covered box to provide additional
isolation from the environment. In order that the box not have to be uncovered
during a measurement data run, electromagnetic shutters were used as much as
possible to block and unblock photon paths and the piezoelectric stage and
camera were computer controlled using data collected by the camera. Because of
these features, all required measurement data were collected before
opto-mechanical instability occurred using only one initial fine alignment. A
data analysis and graphing software tool was developed and used to
automatically process the camera images.

\subsection{Overview}

The essence of this experiment involved comparing the measured weak values of
the operator $\left\vert L2\right\rangle \left\langle L2\right\vert
\equiv\widehat{N}$ for two distinct (data) classes of weak measurements. For
each of these weak measurement classes the pre-selected state prior to the
time of $\widehat{N}$'s measurement was the spatial mode $\left\vert
R1\right\rangle $ and the post-selected state after $\widehat{N}$'s
measurement time was the spatial mode $\left\vert R6\right\rangle $. Also, for
each of these classes the path lengths in the first Mach-Zehnder were arranged
so that photons effectively only emerged from BS2 along path $R4$ in spatial
mode $-\left\vert R4\right\rangle $. Thus, paths $R4$ and $L4$ will be
referred to as the "bright" and "dark" paths, respectively. Arranging the
first Mach-Zehnder in this way corresponded to localizing a photon at the
right slit of a two slit screen prior to its traversing the screen. Weak
measurements of $\widehat{N}$ for both measurement classes were made while the
apparatus was in this configuration - except that a shutter blocked path $L4$
for the second measurement class. Blocking path $L4$ in this manner
corresponded to closing the left slit in a two slit screen while the photon is
localized at the right slit.

If $N_{w,1}$ and $N_{w,2}$ correspond to the weak values of $\widehat{N}$ for
the first and second measurement classes, respectively, then - since $L4$ is a
dark path - it might be expected that blocking path $L4$ should have no effect
upon the weak measurement of $\widehat{N}$ in $L2$, in which case
$N_{w,1}=N_{w,2}$. However, when eq.(\ref{4}) is used to calculate these weak
values it is found that for the first measurement class (which corresponds to
both slits being open) $N_{w,1}=+1$ and for the second measurement class
(which corresponds to closing the left slit) $N_{w,2}=+\frac{1}{2}$. More
specifically, for the first measurement class, forward propagation of the
pre-selected state $\left\vert R1\right\rangle $ and backward propagation of
the post-selected state $\left\vert R6\right\rangle $ through the
interferometer to where $\widehat{N}$ is measured yields the states $\frac
{1}{\sqrt{2}}\left(  i\left\vert L2\right\rangle +\left\vert R2\right\rangle
\right)  $ and $i\left\vert L2\right\rangle $, respectively, so that
\[
N_{w,1}=\frac{\left[  -i\left\langle L2\right\vert \right]  \widehat{N}\left[
\frac{1}{\sqrt{2}}\left(  i\left\vert L2\right\rangle +\left\vert
R2\right\rangle \right)  \right]  }{\left[  -i\left\langle L2\right\vert
\right]  \left[  \frac{1}{\sqrt{2}}\left(  i\left\vert L2\right\rangle
+\left\vert R2\right\rangle \right)  \right]  }=+1
\]
(note that the theoretical weak value of $\left\vert R2\right\rangle
\left\langle R2\right\vert $ is $0$). Similarly, for the second measurement
class - with the dark path $L4$ blocked - forward propagation of the
pre-selected state $\left\vert R1\right\rangle $ and backward propagation of
the post-selected state $\left\vert R6\right\rangle $ through the
interferometer to where $\widehat{N}$ is measured yields the states $\frac
{1}{\sqrt{2}}\left(  i\left\vert L2\right\rangle +\left\vert R2\right\rangle
\right)  $ and $\frac{1}{2}\left(  i\left\vert L2\right\rangle +\left\vert
R2\right\rangle \right)  $, respectively, so that
\[
N_{w,2}=\frac{\left[  \frac{1}{2}\left(  -i\left\langle L2\right\vert
+\left\langle R2\right\vert \right)  \right]  \widehat{N}\left[  \frac
{1}{\sqrt{2}}\left(  i\left\vert L2\right\rangle +\left\vert R2\right\rangle
\right)  \right]  }{\left[  \frac{1}{2}\left(  -i\left\langle L2\right\vert
+\left\langle R2\right\vert \right)  \right]  \left[  \frac{1}{\sqrt{2}%
}\left(  i\left\vert L2\right\rangle +\left\vert R2\right\rangle \right)
\right]  }=+\frac{1}{2}%
\]
(note that the theoretical weak value of $\left\vert R2\right\rangle
\left\langle R2\right\vert $ is also $+\frac{1}{2}$).

Thus, $N_{w,1}\neq N_{w,2}$ so that - similar to the \textit{gedanken}
experiment - weak value theory applied to this experiment predicts that
blocking path $L4$ produces a dramatic observable change in the weak value of
$\widehat{N}$ when there are effectively no photons along path $L4$. Following
\cite{TACKN} and using the two-slit case along with eq.(\ref{1a}) as guides,
$N_{w,1}\neq N_{w,2}$ \emph{has an interpretation as being an effect induced
in the first Mach-Zehnder by the non-local exchange of modular momentum that
results from a change in the potential associated with blocking the dark }%
$L4$\emph{ path in the second Mach-Zehnder.}

A third class of weak measurements of $\widehat{N}$ designated by the weak
value $N_{w,0}$ was used for the purpose of order compliance. For this
measurement class the configuration of the first Mach-Zehnder was the same as
for the other two classes so that forward propagation of the pre-selected
state $\left\vert R1\right\rangle $ through the first Mach-Zehnder yielded the
state $-\left\vert R4\right\rangle $. Here, however, a relative (to the other
two classes) phase shift of $\pi$ $rad$ was introduced into path $R5$ so that
backward propagation of the post-selected state $\left\vert R6\right\rangle $
backwards through the interferometer gives $\left\vert R2\right\rangle $ as
the state where the measurement is made. Again using $\frac{1}{\sqrt{2}%
}\left(  i\left\vert L2\right\rangle +\left\vert R2\right\rangle \right)  $ as
the forward propagated pre-selected state yields the weak value
\[
N_{w,0}=\frac{\left[  \left\langle R2\right\vert \right]  \widehat{N}\left[
\frac{1}{\sqrt{2}}\left(  i\left\vert L2\right\rangle +\left\vert
R2\right\rangle \right)  \right]  }{\left[  \left\langle R2\right\vert
\right]  \left[  \frac{1}{\sqrt{2}}\left(  i\left\vert L2\right\rangle
+\left\vert R2\right\rangle \right)  \right]  }=0
\]
(note that the theoretical weak value of $\left\vert R2\right\rangle
\left\langle R2\right\vert $ is $+1$). This class of measurements served as a
data consistency check by demonstrating that the weak values $N_{w,0}$,
$N_{w,1}$, and $N_{w,2}$ measured by this experiment were compliant with the
theoretical ordering requirement%
\begin{equation}
N_{w,0}<N_{w,2}<N_{w,1}. \label{7}%
\end{equation}

\section{ Results}

In order to experimentally demonstrate this induced $N_{w,1}\neq N_{w,2}$
effect, three sequences of weak measurements of $\widehat{N}$ - one sequence
for each of the $N_{w,0}$, $N_{w,1}$, and $N_{w,2}$ measurement classes - were
generated following the configuration prescriptions outlined in the above
overview of the experiment. Different interaction strength ($\gamma$) values
were produced for each sequence by varying the M1 position via controlling
that of the piezoelectric stage. The photon beam intensity served as the
measurement pointer for the apparatus and its image was recorded by the
machine vision camera for each M1 position used in the measurement sequences.
As indicated on figure 1, the associated movement of the pointer in the image
plane was horizontal (i.e. in the plane of the apparatus). For each M1
position $x$, the analysis software tool used the associated pointer image to
locate the pointer position as the intensity averaged horizontal pixel number
$\overline{y}$. Each such measurement was represented as the pair $\left(
x,\overline{y}\right)  $. Let $S_{i}$ be the set of such measurement pairs for
the $N_{w,i}$ measurement class, $i=0,1,2$.

To calibrate the experimental data, a fourth sequence of measurements was made
to relate M1 positions to pointer pixel positions. Here, paths $L3$ and $R4$
were blocked by shutters and a sequence of M1 positions were used to sweep the
beam emerging along path $R6$ across the image plane of the camera. As was the
case for the previous sequences of measurements, the beam's intensity averaged
horizontal pixel number was determined from each M1 position image and
represented as an ordered pair $\left(  x,\overline{y}\right)  $. Let $S_{3}$
be the set of these ordered pairs of calibration measurements.

Fourteen M1 positions equally spaced over a $1300$ $\mu m$ range were used to
generate fourteen ordered pairs of measurements in each set $S_{k}$, $k\in
K\equiv\left\{  0,1,2,3\right\}  $. These M1 positions were identical for each
of the four measurement sequences (i.e. for every $\left(  x,\overline
{y}\right)  \in S_{k}$, $k\in K$, there is exactly one $\left(  x^{\prime
},\overline{y}^{\prime}\right)  \in S_{j}$, $j\in K-\left\{  k\right\}  $,
such that $x=x^{\prime}$). Examination of the measurement pairs in $S_{1}$ and
$S_{2}$ revealed the existence of a data crossing point located between the
middle two M1 positions $x_{7}$ and $x_{8}$. The pair $\left(  x_{0}%
,y_{0}\right)  $ defined by the intersection of the line containing the middle
two measurement pairs in $S_{1}$ with that containing the middle two
measurement pairs in $S_{2}$ was selected as the estimate of this crossing point.

Recall from eq.(\ref{3}) that in the weak measurement regime defined by
inequalities (\ref{5}) the ordinates in each data pair in the sets $S_{i}$
effectively record the measured quantity $\gamma N_{w,i}$, $i=0,1,2$. Thus, at
the crossing point the condition $\gamma N_{w,1}=\gamma N_{w,2}$ must hold
true. Since $N_{w,1}=+1$ and $N_{w,2}=+\frac{1}{2}$, this condition can only
be satisfied if $\gamma=0$. This identified $\left(  x_{0},y_{0}\right)  $ as
the point where the interaction strength $\gamma$ vanishes and defined it as
the origin of the Cartesian reference frame $\mathcal{F}$ which has as its
abscissa axis M1 displacements in $\mu m$ referenced to $x_{0}$ and as its
ordinate axis pointer pixel displacements referenced to $y_{0}$. Let $\left(
x^{\prime},\overline{y}^{\prime}\right)  \in S_{k}^{\prime}$ be $\left(
x,\overline{y}\right)  \in S_{k}$, $k\in\left\{  0,1,2\right\}  $, transformed
into $\mathcal{F}$ according to $x^{\prime}=x-x_{0}$ and $\overline{y}%
^{\prime}=\overline{y}-y_{0}$.

As anticipated - the calibration measurement pairs in $S_{3}$ were linear. The
associated slope which relates pointer pixel positions to M1 positions in $\mu
m$ was $-0.198$. This slope defined the calibration line $\overline{y}%
^{\prime}=-0.198x^{\prime}$ in $\mathcal{F}$. Multiplying the slope of this
equation by the pixel size $7.4$ $\mu m$ (the camera rated distance between
consecutive pixels) yielded the equation $\gamma\left(  x^{\prime}\right)
=-1.5x^{\prime}$\ in which both $\gamma$ and $x^{\prime}$ are in $\mu m$. The
ordinate $\overline{y}^{\prime}$ is relabeled as $\gamma\left(  x^{\prime
}\right)  $ in this equation because it now directly relates the interaction
strengths of measurements to the displacement of M1 (inspection of the
argument of the operator $\widehat{S}$ in eq.(\ref{2b}) reveals that $\gamma$
is a distance since $N_{w,i}$ is a dimensionless quantity). Thus - for this
experiment \ - the "ideal" pointer displacements $\rho_{i}\left(  x^{\prime
}\right)  \equiv\gamma\left(  x^{\prime}\right)  N_{w,i}$ in $\mu m$ as
functions of M1 displacements in $\mu m$ and $N_{w,i}$ values are represented
by the lines%
\begin{equation}
\rho_{i}\left(  x^{\prime}\right)  =-1.5x^{\prime}N_{w,i}\text{, }i=0,1,2.
\label{6}%
\end{equation}

This result is useful for estimating the boundaries of the weak measurement
regime for this experiment in terms of $x^{\prime}$. Since $\widehat{N}$ is a
projection operator then $\widehat{N}^{m}=\widehat{N}$, $m\geq1$, so that
$\left(  N^{m}\right)  _{w,i}=N_{w,i}$ and inequalities (\ref{5}) become
$\Delta p\ll\frac{\hbar}{\gamma N_{w,i}}$, $i\neq0$, and $\Delta p\ll
\frac{\hbar}{\gamma}$. Application of the uncertainty relation $\Delta
q\cdot\Delta p\geq\frac{\hbar}{2}$ yields $\gamma\ll\frac{2\Delta q}{N_{w,i}}%
$, $i\neq0$, and $\gamma\ll2\Delta q$. Both of these inequalities are
satisfied by $\gamma\ll2\Delta q$ when $i=1,2$. Using the pinhole diameter as
the uncertainty in a photon's tranverse position, i.e. $\Delta q\approx150$
$\mu m$, defines $\left\vert \gamma\right\vert \ll300$ $\mu m$ as the
estimated weak measurement regime for the interaction strength ($\left\vert
\gamma\right\vert $ is used since in this experiment $\gamma$ can be a
positive or a negative distance). Using this range in eq.(\ref{6}) with
$N_{w,1}=+1$ gives $\rho_{1}\left(  x^{\prime}\right)  =\gamma\left(
x^{\prime}\right)  $ and yields
\begin{equation}
\left\vert x^{\prime}\right\vert \ll200\text{ }\mu m \label{8}%
\end{equation}
as the estimated weak measurement regime for M1 displacement.

A plot of the measurement pairs in sets $S_{i}^{\prime}$, $i=0,1,2$ is
presented in figure 2. Here the ordinate of each measurement pair has been
scaled by the pixel distance of $7.4$ $\mu m$ in order to express the pointer
displacements in $\mu m$. Also shown as dashed lines are graphs of the three
ideal pointer displacement lines $\rho_{i}\left(  x^{\prime}\right)  $,
$i=0,1,2$, given by eq.(\ref{6}) and as a boxed region the estimated weak
measurement regime defined by inequality (\ref{8}). Inspection of figure 2
(where $\gamma N_{w,i}$ data points are labeled "$\gamma N$ class $i$" and
$\rho_{i}$ is labeled "$\rho$ class $i$") reveals good agreement within (and
slightly outside) the weak measurement regime between the measured pointer
displacements $\gamma N_{w,1}$ (corresponding to the measurement pairs in set
$S_{1}^{\prime}$) and $\rho_{1}$ and between the measured pointer
displacements $\gamma N_{w,2}$ (corresponding to the measurement pairs in set
$S_{2}^{\prime}$) and $\rho_{2}$. It is also clear that - except at
$x_{8}^{\prime}$ - the measured quantities within the weak measurement regime
are compliant with the theoretical ordering requirement (\ref{7}). It is noted
that the $\sim75$ $\mu m-100$ $\mu m$ offsets of the measured pointer
displacements $\gamma N_{w,0}$ (corresponding to the measurement pairs in set
$S_{0}^{\prime}$) from $\rho_{0}$ in the weak measurement regime are likely
due to complicated intensity profile inversions introduced by the phase window
during this sequence of measurements. Interestingly, if these offsets are
treated as a constant bias, then removal of the bias from the measurement
pairs in $S_{0}^{\prime}$ not only produces complete compliance with (\ref{7})
in the weak measurement regime - but it also provides more overall symmetry in
the data, as well as good agreement between the measured pointer displacements
$\gamma N_{w,0}$ and $\rho_{0}$ in the weak measurement regime.

As expected, the further the M1 displacement is outside the weak measurement
regime the "stronger" the measurement becomes and the greater the discrepancy
between the $S_{0}^{\prime}$ data and $\rho_{0}$ and between the
$S_{1}^{\prime}$ data and $\rho_{1}$. However, except for the data asymmetry
associated with negative M1 displacements (likely introduced by the
complicated optical properties of the apparatus), the agreement between the
$S_{2}^{\prime}$ data and $\rho_{2}$ remains good over the entire range of M1
displacements while the $S_{0}^{\prime}$ and $S_{1}^{\prime}$ data converge to
$\rho_{2}$. This feature in the data is completely consistent with the fact
that in the limit of "strong collapsing" measurements, the measurement pointer
is displaced by $\gamma\left\langle N\right\rangle =\frac{1}{2}\gamma$ since%
\[
\left\langle N\right\rangle =\frac{1}{\sqrt{2}}\left[  -i\left\langle
L2\right\vert +\left\langle R2\right\vert \right]  \widehat{N}\left[
i\left\vert L2\right\rangle +\left\vert R2\right\rangle \right]  \frac
{1}{\sqrt{2}}=+\frac{1}{2}%
\]
(refer to the discussion surrounding eq.(\ref{2})).

\section{Concluding Remarks}

This experiment used weak measurements of pre- and post-selected ensembles of
photons in a twin Mach-Zehnder interferometer to observe an effect
theoretically predicted to be induced in the first Mach-Zehnder by the
non-local exchange of modular momentum produced by blocking the dark path in
the second Mach-Zehnder (it is intended that a second "follow up" paper be
written which will detail the novel aspects of the apparatus and techniques
used in this experiment). This effect is manifested as a dramatic change in
the associated weak values. The attendant weak values measured by this
experiment changed in complete accordance with the theoretical predictions.
Consequently, the results of this experiment support both the existence of
such an effect and the authenticity of dynamical non-locality as its cause.

Before closing, it is noted that - although this experiment was specifically
designed for the purpose of confirming or denying the $N_{w,1}\neq N_{w,2}$
effect - it was observed that - for the weakest measurements with abscissa
$x_{8}\simeq37$ $\mu m$ - the ratio of the number of camera pixels excited by
the associated $S_{2}^{\prime}$ measurement to that excited by the associated
$S_{1}^{\prime}$ measurement was $0.6$. The drop in this excitation ratio was
$4$ to $5$ times greater than expected based upon the alignment contrast
ratios for the apparatus. This informal observation provides additional
credence to dynamical non-locality as inducing the $N_{w,1}\neq N_{w,2}$
effect and suggests a future experiment that could further examine dynamical
non-locality from this perspective.

\begin{acknowledgments}
The authors thank Yakir Aharonov and Jeff Tollaksen for suggesting this
experiment; John Gray and James Troupe for constructive technical discussions;
and David Niemi for his efforts in the instrumentation of this experiment.
Special thanks are given to Susan Hudson, Electromagnetic and Sensor Systems
Department Head, for her commitment to this research. This work was supported
in part by a grant from the NSWCDD ILIR program sponsored by the Office of
Naval Research.
\end{acknowledgments}

\bigskip%

\begin{equation}%
\begin{array}
[c]{cccccccccccccccccc}
&  &  &  &  &  &  &  &  &  &  &  &  &  &  &  &  & \\
&  &  &  &  &  &  &  &  &  &  &  &  &  &  &  &  & \\
&  &  &  &  &  &  &  &  & \cdot & \cdots & \cdots & \cdots & \cdots & \cdots &
\cdots & Phase & \\
&  &  &  &  &  &  &  &  & \vdots &  &  &  &  &  &  & window & \\
&  &  &  &  &  &  &  &  & \vdots &  &  & \uparrow &  &  &  &  & \\
&  &  &  &  &  &  &  &  & \vdots &  &  & L6 &  &  &  &  & \\
&  &  &  &  &  &  &  &  & \vdots &  &  & \uparrow &  &  &  &  & \\
&  &  &  &  &  &  & M3 & \diagup & \cdots & R5 & \cdots & \diagup &
\longrightarrow & R6 & \longrightarrow & \updownarrow\text{ }Camera & \\
&  &  &  &  &  &  & \diagdown & \vdots &  &  & BS3 & \uparrow &  &  &  &
with\text{ }directions & \\
&  &  &  &  &  &  &  & L4 &  &  &  & L5 &  &  &  & for\text{ }pointer & \\
&  &  & \nwarrow\searrow &  &  &  &  & \vdots & \diagdown &  &  & \uparrow &
&  &  & movement & \\
&  &  & M1 & \diagup & \longrightarrow & R3 & \longrightarrow & \diagup &
\longrightarrow & R4 & \longrightarrow & \diagup & M4 &  &  &  & \\
&  &  &  & \uparrow &  &  & BS2 & \uparrow &  &  & \diagdown & \cdots & \cdots
& \cdots & \cdots & Metaphorical & \\
&  &  &  & L2 &  &  &  & L3 &  &  &  &  &  &  &  & two-slits & \\
&  &  &  & \uparrow &  &  &  & \uparrow &  &  &  &  &  &  &  &  & \\
Laser & \longrightarrow & R1 & \longrightarrow & \diagup & \longrightarrow &
R2 & \longrightarrow & \diagup & M2 &  &  &  &  &  &  &  & \\
beam &  &  & BS1 &  &  &  &  &  &  &  &  &  &  &  &  &  & \\
&  &  &  &  &  &  &  &  &  &  &  &  &  &  &  &  &
\end{array}
\tag{Figure 1. Apparatus, best available diagram for electronic publishing}%
\end{equation}

\bigskip

\bigskip

\bigskip

\bigskip$%
{\includegraphics[
natheight=3.850100in,
natwidth=5.000400in,
height=5.489in,
width=7.1174in
]%
{Results.bmp}%
}%
$

\bigskip

\bigskip

\bigskip

\bigskip

\bigskip

\bigskip

\bigskip

\bigskip

\bigskip

\bigskip

$\qquad\qquad$

\begin{center}

\end{center}

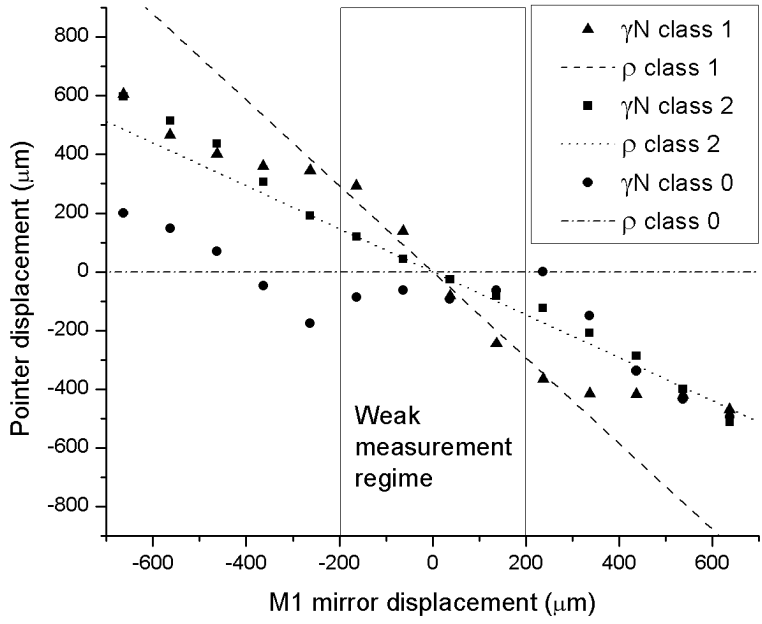
\end{document}